\documentclass[12pt]{article}

\usepackage{amssymb}
\usepackage{amsmath}
\usepackage{amsfonts}

\renewcommand{\baselinestretch}{1.2}

\newcommand{\R}{\mathbb{R}}
\newcommand{\C}{\mathbb{C}}
\newcommand{\Z}{\mathbb{Z}}

\newcommand{\be}{\begin{equation}}
\newcommand{\bea}{\begin{eqnarray}}
\newcommand{\eea}{\end{eqnarray}}
\newcommand{\nn}{\nonumber}
\newcommand{\kt}{\rangle}
\newcommand{\br}{\langle}

\newcommand{\ed}{\end{document}}
\newcommand{\bbr}{\br\!\br}
\newcommand{\kkt}{\kt\!\kt}

\begin{document}

\title{A Physical Realization of the Generalized
$PT$-, $C$-, and $CPT$-Symmetries and the Position Operator for
Klein-Gordon Fields}
\author{\\
Ali Mostafazadeh\thanks{E-mail address: amostafazadeh@ku.edu.tr}\\
\\ Department of Mathematics, Ko\c{c} University,\\
Rumelifeneri Yolu, 34450 Sariyer,\\
Istanbul, Turkey}
\date{ }
\maketitle
\begin{abstract}
Generalized parity (${\cal P}$), time-reversal (${\cal T}$), and
charge-conjugation (${\cal C}$) operators were initially defined
in the study of the pseudo-Hermitian Hamiltonians. We construct a
concrete realization of these operators for Klein-Gordon fields
and show that in this realization ${\cal PT}$ and ${\cal C}$
operators respectively correspond to the ordinary time-reversal
and charge-grading operations. Furthermore, we present a complete
description of the quantum mechanics of Klein-Gordon fields that
is based on the construction of a Hilbert space with a
relativistically invariant, positive-definite, and conserved inner
product. In particular we offer a natural construction of a
position operator and the corresponding localized and coherent
states. The restriction of this position operator to the
positive-frequency fields coincides with the Newton-Wigner
operator. Our approach does not rely on the conventional
restriction to positive-frequency fields. Yet it provides a
consistent quantum mechanical description of Klein-Gordon fields
with a genuine probabilistic interpretation.
\end{abstract}
%PACS numbers: 98.80.Qc, 04.60.-m, 03.65.Pm
%\vspace{2mm}

%\baselineskip=24pt

\section{Introduction}

Since the formulation of quantum mechanics (QM) in the 1920s,
there has been an ongoing search for its generalizations. Among
the early examples of such a generalization is Pauli's attempts to
formulate a quantum mechanical theory based on a vector space with
an indefinite inner product \cite{pauli}. This was originally
motivated by Dirac's observation that such theories would ease
handling the infinities arising in quantum electrodynamics
\cite{dirac-1942}. QM with an indefinite metric has been
thoroughly studied in the 1960s \cite{sudarshan-1961}. Another
more recent attempt at a generalization of QM is so-called
$PT$-Symmetric QM \cite{bender}.

The interest in $PT$-symmetric QM relies on an appealing idea due
to Bender and a set of rather surprising spectral properties of
certain non-Hermitian but $PT$-symmetric Hamiltonians of the
standard form: $H=p^2+V(x)$.\footnote{Here $P$ and $T$ are
respectively the parity and time-reversal operators defined in the
space of all complex-valued functions $\psi:\R\to\C$ by
$(P\psi)(x):=\psi(-x)$ and $(T\psi)(x):=\psi(x)^*$. The letter
relation defines $T$ as an antilinear operator whose action should
not be viewed as a change of sign of $t$.} Bender's idea concerned
the possibility that one might be able to generalize the standard
local quantum field theory by replacing the postulate of the
Hermiticity of the Hamiltonian by the supposedly more general
requirement of its $CPT$-symmetry. $(0+1)$-dimensional examples of
such a field theory are provided by the $PT$-symmetric QM in which
one allows the Hamiltonian to be non-Hermitian but requires that
it commutes with $PT$. The study of $PT$-symmetric quantum systems
gained some valid ground once it became clear that the spectrum of
some simple $PT$-symmetric Hamiltonians, such as $H=p^2+ix^3$, was
entirely real and positive, \cite{dorey}.

During the past seven years, there have appeared dozens of
publications on $PT$-symmetric QM. Most of these examined specific
toy models \cite{models}, but some attempted to address the more
fundamental issues such as that of the nature of the space of
state vectors \cite{m1} -- \cite{p7}. Among the latter is a series
of articles \cite{p1} -- \cite{p7} by the present author that aim
at providing a mathematically sound framework for describing
$PT$-symmetric QM and its physical interpretation \cite{jpa04c}.
The results reported in these articles stem from the rather simple
observation that, similarly to the Hermitian Hamiltonians, the
$PT$-symmetric Hamiltonians constitute a subclass of the so-called
pseudo-Hermitian Hamiltonians. The latter are linear operators $H$
acting in a Hilbert space ${\cal H}'$ --- which is generally
different from the physical Hilbert space --- and satisfying
    \be
    H^\dagger=\eta H\eta^{-1}
    \label{p-h}
    \end{equation}
for some invertible Hermitian operator $\eta:{\cal H}'\to{\cal
H}'$. Here $H^\dagger$ denotes the adjoint of $H$ defined using
the inner product $\br\cdot,\cdot\kt$ of the Hilbert space ${\cal
H}'$. The operator $\eta$ defines a possibly indefinite inner
product, namely
    \be
    \bbr\cdot,\cdot\kkt_\eta:=\br\cdot,\eta\cdot\kt.
    \label{indef-inn}
    \end{equation}
As a consequence, it is sometimes called a `metric operator'. The
main property of (\ref{indef-inn}) is that $H$ is Hermitian with
respect to this inner product, i.e.,
$\bbr\cdot,H\cdot\kkt_\eta=\bbr H\cdot,\cdot\kkt_\eta$.

For a given pseudo-Hermitian operator $H$, the metric operator
$\eta$ is not unique. If a particular choice for $\eta$ is made,
then $H$ is called $\eta$-pseudo-Hermitian. The standard
$PT$-symmetric models studied in the literature are
$P$-pseudo-Hermitian \cite{p1}. See also \cite{p5,z-ahmed}.

It turns out that, under quite general conditions, the
pseudo-Hermiticity of $H$ is equivalent to the existence of
antilinear symmetries of $H$, $PT$-symmetry being the primary
example \cite{p3,s1,ss2}. Furthermore, pseudo-Hermiticity of $H$
is equivalent to the condition that either the spectrum of $H$ is
real or its complex eigenvalues come in complex-conjugate pairs.
The study of pseudo-Hermitian operators also provides a
characterization of the non-Hermitian (diagonalizable)
Hamiltonians that have a real spectrum. These correspond to the
subclass of pseudo-Hermitian Hamiltonians that are
$\eta_+$-pseudo-Hermitian with respect to some
positive-definite\footnote{A positive-definite operator is an
invertible positive operator. Alternatively, it is a self-adjoint
operator with a strictly positive spectrum.} metric operator
$\eta_+$. In this case $H$ is Hermitian with respect to a genuine
positive-definite inner product\footnote{The positive-definiteness
of an inner product $\bbr\cdot,\cdot\kkt$ means that
$\bbr\psi,\psi\kkt\in\R^+$ if and only if $\psi\neq 0$.}, namely
    \be
    \bbr\cdot,\cdot\kkt_{\eta_+}:=\br\cdot,\eta_+\cdot\kt.
    \label{inn}
    \end{equation}
If $H$ does not have a real spectrum or if it is not
diagonalizable, the set of all metric operators $\eta$
satisfying~(\ref{p-h}) does not include any positive-definite
elements and such a positive-definite inner product does not exist
\cite{p6,ss2}.

It can also be shown that whenever $H$ is a diagonalizable
operator with a real spectrum, then it can be mapped to a
Hermitian operator via a similarity transformation \cite{p2},
i.e., it is quasi-Hermitian \cite{quasi}. Quasi-Hermitian
Hamiltonians form a proper subset of pseudo-Hermitian Hamiltonians
which allow for the formulation of a probabilistic quantum theory.
This is done by identifying the physical Hilbert space ${\cal H}$
with the (Cauchy completion \cite{reed-simon} of the) span of the
eigenvectors of $H$ (i.e., the underlying vector space associated
with the Hilbert space ${\cal H}'$) endowed with the inner product
$\bbr\cdot,\cdot\kkt_{\eta_+}$, \cite{reed-simon,k}. By
construction, $H$ will be a Hermitian (densely-defined) operator
acting in ${\cal H}$. Similarly the observables of the theory are
the Hermitian operators acting in ${\cal H}$. For concrete
examples see \cite{jpa04c}.

The reality of the spectrum of $H$ is also equivalent to the
presence of an exact antilinear symmetry of the system. Therefore,
a diagonalizable Hamiltonian having an exact antilinear symmetry,
e.g., $PT$-symmetry, is quasi-Hermitian. In particular, it may be
mapped to a Hermitian Hamiltonian $H'$ via a similarity
transformation\footnote{This applies for the cases that $H$ has
time-independent eigenvectors. Throughout this article we will
only consider time-independent Hamiltonians.} $H\to H'={\cal
U}H{\cal U}^{-1}$ where the operator ${\cal U}$ satisfies $\bbr
\cdot,\cdot\kkt_{\eta_+}=\br{\cal U}\cdot,{\cal U}\cdot\kt$, i.e.,
${\cal U}$ is a unitary operator mapping ${\cal H}$ onto ${\cal
H}'$. This shows that indeed the quantum system defined by the
Hamiltonian $H$ and the Hilbert space ${\cal H}$ may as well be
described by the Hermitian Hamiltonian $H'$ and the original
Hilbert space ${\cal H}'$, \cite{p10}.

In an attempt at devising a probabilistic interpretation for
quantum systems with Hamiltonians having exact $PT$-symmetry,
Bender and his collaborators have introduced a generic symmetry of
these Hamiltonians that they term as the `charge-conjugation'
symmetry \cite{bender3}. Using the generator $C$ of this symmetry,
they were able to introduce a positive-definite inner product that
they called the $CPT$-inner product. As shown in \cite{p7,jpa05a}
the $CPT$-inner product is just a particular example of the inner
product (\ref{inn}) that was originally constructed in
Refs.~\cite{p2,p3,p4}.

The description of the `charge conjugation' operator $C$ of the
$PT$-symmetric QM \cite{bender3} provided by the theory of
pseudo-Hermitian Hamiltonians and the fact that general
pseudo-Hermitian Hamiltonians have generic antilinear symmetries
\cite{p3,s1,ss2} raise the natural question whether one could
associate to a general pseudo-Hermitian Hamiltonian a linear
symmetry generator ${\cal C}$ and an antilinear symmetry generator
${\cal PT}$ that would respectively generalize $C$ and $PT$. This
question is answered in \cite{p7}.\footnote{See also
\cite{z-ahmed2} where the same question has been considered.} It
turns out that for a given diagonalizable pseudo-Hermitian
Hamiltonian $H$ one may introduce generalized parity $({\cal P})$,
time-reversal $({\cal T})$ and charge-conjugation $({\cal C})$
operators and establish the ${\cal PT}$-, ${\cal C}$-, and ${\cal
CPT}$-symmetries of $H$, \cite{p7}. It must be noted that the use
of the term `charge-conjugation' in the above discussions solely
rests on the fact that similarly to the ordinary
charge-conjugation operator of relativistic QM, ${\cal C}$ is a
Hermitian involution, i.e., ${\cal C}^\dagger={\cal C}={\cal
C}^{-1}$. It is important to note that because ${\cal C}$ is a
linear operator, it is actually a $\Z_2$-grading operator for the
Hilbert space \cite{npb-01}.

The question whether ${\cal C}$ has anything to do with the
well-known charge-conjugation operation of relativistic QM is the
principal goal of the present paper. We will achieve this goal by
constructing a concrete realization of the abstract notions of
generalized parity, time-reversal, and charge-conjugation
operations for Klein-Gordon fields. This will in turn allow us to
offer an application of the theory of pseudo-Hermitian operators
in addressing one of the oldest problems of modern physics
concerning relativistic position operators and localized states
\cite{newton-wigner}.

The problem of constructing a consistent quantum mechanical
formulation of Klein-Gordon fields has been the focus of attention
since the early days of QM. There are numerous publications on the
subject most of which are based on the restriction to
positive-frequency fields \cite{FV,Greiner,GS}. A notable
exception is a series of papers by Gitman and his collaborators
\cite{gt1,gt2} which were brought to our attention while we were
writing up the present paper.\footnote{See also \cite{saa}.} In
\cite{gt1,gt2} the authors base their investigation on a careful
treatment of the underlying classical system and perform a
constraint canonical quantization of this system. In particular,
they identify a discrete classical degree of freedom $\zeta$ which
takes  the values $+1$ and $-1$. Quantization of $\zeta$ leads to
a quantum observable $\hat\zeta$ which satisfies $\hat\zeta^2=1$
and plays a key role in the approach of \cite{gt1,gt2}. We will
elaborate on the relationship between this approach and ours in
Section~4. Here we suffice to mention that $\zeta$ arises in our
treatment as the classical counterpart of the generalized
charge-conjugation operator ${\cal C}$.

The article is organized as follows. In Section~2, we present a
two-component formulation of the Klein-Gordon equation, establish
the pseudo-Hermiticity and quasi-Hermiticity of the corresponding
two-component matrix Hamiltonian, construct a positive-definite
metric operator and the corresponding inner product, discuss the
relationship between the one- and two-component fields and review
some of the relevant results reported in \cite{cqg,p9}. In
Section~3, we compute ${\cal P}$, ${\cal T}$, and ${\cal C}$ for
the two-component Klein-Gordon fields and elaborate on the ${\cal
PT}$-, ${\cal C}$- and ${\cal CPT}$-symmetries of the
corresponding Hamiltonian. We then define and study generalized
parity, time-reversal, and charge-conjugation operators for the
ordinary single-component Klein-Gordon fields. In Section~4, we
introduce and study position wave functions for the Klein-Gordon
fields. In Section~5, we discuss the position and momentum
operators and the relativistic localized and coherent states.
Finally in Section~6, we present a summary of our main results and
our concluding remarks.

Throughout this article we shall use a unit system in which $c=1$.

\section{Pseudo-Hermiticity and Klein-Gordon Fields}

Consider the Klein-Gordon equation
    \be
    [\partial_t^2-\nabla^2+\mu^2]\psi(t,\vec x)=0,
    \label{kg}
    \end{equation}
where $\mu:=m/\hbar$ and $m$ is the mass of the field. Suppose
that, for all $t\in\R$, $\psi(t,\vec x)$ is square-integrable:
$\int_{\R^3}d^3x~|\psi(t,\vec x)|^2<\infty$. Then we can
express~(\ref{kg}) as a `dynamical equation' in the Hilbert space
$L^2(\R^3)$, namely
    \be
    [\partial_t^2+D]\psi(t)=0,
    \label{kgt}
    \end{equation}
where for all $t\in\R$ the function $\psi(t):\R^3\to\C$, defined
by $\psi(t)(\vec x):=\psi(t,\vec x)$, belongs to $L^2(\R^3)$, and
$D:L^2(\R^3)\to L^2(\R^3)$ is the Hermitian operator
    \be
    [D\psi(t)](\vec x):=[-\nabla^2+\mu^2]\psi(t,\vec x),~~~~~~~~~~
    \forall t\in\R,~\forall\vec x\in\R^3.
    \label{kg-D}
    \end{equation}

It is well-known \cite{foldy,FV,Greiner,jpa-98} that one can
express the Klein-Gordon equation~(\ref{kgt}) as the two-component
Schr\"odinger equation:
    \be
    i\hbar \frac{d}{dt}\Psi(t)=H\Psi(t),
    \label{2-comp-sch-eq}
    \end{equation}
where for all $t\in\R$
    \bea
    \Psi(t)&:=&\left(\begin{array}{c}
    \psi(t)+i\lambda\dot\psi(t)\\
    \psi(t)-i\lambda\dot\psi(t)\end{array}\right),
    \label{2-comp-psi}\\
    H&:=&\frac{\hbar}{2}\,\left(\begin{array}{cc}
    \lambda D+\lambda^{-1}& \lambda D-\lambda^{-1}\\
    -\lambda D+\lambda^{-1}& -\lambda D-\lambda^{-1}
    \end{array}\right),
    \label{2-comp-H}
    \eea
a dot denotes a $t$-derivative, and $\lambda\in\R-\{0\}$ is an
arbitrary constant having the dimension of length.

The two-component vectors $\Psi(t)$ belong to
    \be
    {\cal H}':=L^2(\R^3)\oplus L^2(\R^3),
    \label{prime}
    \end{equation}
and the Hamiltonian~$H$ may be viewed as acting in ${\cal H}'$.
One can easily check that $H$ is not Hermitian with respect to the
inner product of ${\cal H}'$, but it satisfies $H^\dagger=\sigma_3
H\sigma_3^{-1}$, where $\sigma_3:=${\tiny$\left(\begin{array}{cc}
    1 & 0 \\
    0 & -1\end{array}\right)$},
\cite{FV, Greiner}. Hence $H$ is $\sigma_3$-pseudo-Hermitian
\cite{p1}. Furthermore, because $D$ is positive-definite, $H$ is a
diagonalizable operator with a real spectrum. This in turn implies
that it is quasi-Hermitian \cite{p2,quasi}. According to
\cite{p2}, $H$ is $\eta_+$-pseudo-Hermitian for a
positive-definite metric operator $\eta_+$. Equivalently, it is
Hermitian with respect to the corresponding positive-definite
inner product, namely (\ref{inn}). The construction of $\eta_+$
requires the solution of the eigenvalue problem for $H$ and
$H^\dagger$.

The eigenvalues and a set of eigenvectors of $H$ are given by
    \bea
    E_{\epsilon,k}&:=&\epsilon\,\hbar\,\omega_k,
    \label{eigenvalue}\\
    \Psi_{\epsilon,\vec k}&:=&
    \frac{1}{2}\,\left(\begin{array}{cc}
    r_k^{-1}+\epsilon r_k\\
    r_k^{-1}-\epsilon r_k\end{array}\right)\,\phi_{\vec k},
    \label{eigenvector-H}
    \eea
where $\epsilon=\pm 1$, $k:=|\vec k|$, $\vec k\in\R^3$,
$\omega_k:=\sqrt{k^2+\mu^2}$, $r_k:=\sqrt{\lambda\,\omega_k}$,
$\phi_{\vec k}$ are defined by $\phi_{\vec k}(\vec
x):=(2\pi)^{-3/2}\;e^{i\vec k\cdot\vec x}=\br\vec x|\vec k\kt$,
and $\br\cdot|\cdot\kt$ denotes the inner product of $L^2(\R^3)$.
The eigenvectors $\Psi_{\epsilon,\vec k}$ together with
    \be
    \Phi_{\epsilon,\vec k}:=
    \frac{1}{2}\,\left(\begin{array}{cc}
    r_k+\epsilon r_k^{-1}\\
    r_k-\epsilon r_k^{-1}\end{array}\right)\,\phi_{\vec k}
    \label{eigenvector-H-dagger}
    \end{equation}
form a complete biorthonormal system for the Hilbert space. This
means that
    \[\br\Psi_{\epsilon,\vec k},\Phi_{\epsilon',\vec k'}\kt=
    \delta_{\epsilon,\epsilon'}\;\delta^3(\vec k-\vec k')
    ,~~~~~~~~~~~~~~~~~~
    \sum_{\epsilon=\pm}\int_{\R^3}d^3k~ |\Psi_{\epsilon,\vec k}\kt
    \br\Phi_{\epsilon,\vec k}|=1,\]
where $\br\cdot,\cdot\kt$ stands for the inner product of ${\cal
H}'$, and for $\xi,\zeta\in{\cal H}'$, $|\xi\kt\br\zeta|$ is the
operator defined by $|\xi\kt\br\zeta|\chi:=\br\zeta,\chi\kt\xi$,
for all $\chi\in{\cal H}'$. Similarly, one can check that indeed
$\Phi_{\epsilon,\vec k}$ are eigenvectors of $H^\dagger$ with the
same eigenvalues, $H^\dagger \Phi_{\epsilon,\vec
k}=E_{\epsilon,k}\,\Phi_{\epsilon,\vec k}$, and that $H$ has the
following spectral resolution
    \be
    H=\sum_{\epsilon=\pm}\int_{\R^3}d^3k~
    E_{\epsilon,k}|\Psi_{\epsilon,\vec k}\kt
    \br\Phi_{\epsilon,\vec k}|.
    \label{spec-reso}
    \end{equation}

The positive-definite metric operator associated with the choice
$\{\Psi_{\epsilon,\vec k},\Phi_{\epsilon,\vec k}\}$ for a
biorthonormal system for ${\cal H}'$ has the from
\cite{p1,p4,p7,cqg}
    \be
    \eta_+=\sum_{\epsilon=\pm}\int_{\R^3}d^3k~
    |\Phi_{\epsilon,\vec k}\kt\br\Phi_{\epsilon,\vec k}|
    =\frac{1}{2}\left(\begin{array}{cc}
    X^2+X^{-2} & X^2-X^{-2}\\
    X^2-X^{-2} & X^2+X^{-2}\end{array}\right):{\cal H}'\to{\cal H}',
    \label{eta+}
    \end{equation}
where
    \be
    X:=\sqrt\lambda~D^{1/4}.
    \label{X}
    \end{equation}
Note that here and throughout this paper we use the spectral
resolution of $D$ to define its powers:
$D^\nu:=\int_{\R^3}d^3k~(k^2+\mu^2)^{\nu}~|\vec
    k\kt\br\vec k|$ for all $\nu\in\R$.

Having obtained the positive-definite metric operator $\eta_+$, we
can compute the form of the corresponding inner product. For all
$\xi,\zeta\in{\cal H}'$, let $\xi^a,\zeta^a\in L^2(\R^3)$ be such
that
    \be
    \xi=\left(\begin{array}{cc}\xi^1\\ \xi^2\end{array}\right),
    ~~~~~~~~~~
    \zeta=\left(\begin{array}{cc}\zeta^1\\
    \zeta^2\end{array}\right).
    \label{xi-zeta}
    \end{equation}
Then in view of (\ref{eta+}),
    \be
    \bbr\xi,\zeta\kkt_{\eta_+}=\frac{1}{2}\,\left[
    \br \xi_+|X^2\zeta_+\kt+\br \xi_-|X^{-2}\zeta_-\kt\right],
    \label{inn-eta+}
    \end{equation}
where we have defined $\xi_\pm:=\xi^1\pm\xi^2$ and
$\zeta_\pm:=\zeta^1\pm\zeta^2$. If we view ${\cal H}'$ as a
complex vector space and endow it with the inner product
(\ref{inn-eta+}), we obtain a new inner product space whose Cauchy
completion yields another Hilbert space which we denote by ${\cal
K}$.

Next, let ${\cal D}\subset L^2(\R^3)$ be the domain of the
operator $D$, and ${\cal V}$ denote the complex vector space of
solutions of the Klein-Gordon equation~(\ref{kgt}), namely
    \be
    {\cal V}:=\left\{\psi:\R\to {\cal D}~|~
    [\partial_t^2+D]\psi(t)=0,~~\forall t\in\R\right\}.
    \label{V}
    \end{equation}
Then, for all $t\in\R$, one can use the map $U_{t}:{\cal
V}\to{\cal H}'$, defined by
    \be
    U_{t}\psi:=\frac{1}{2\sqrt{\lambda\mu}}\:\Psi(t),~~~~~~~~
    \forall \psi\in{\cal V},
    \label{U-zero}
    \end{equation}
where $\Psi(t)$ is given in (\ref{2-comp-psi}), to endow ${\cal
V}$ with the positive-definite inner product
    \be
    (\psi_1,\psi_2):=\bbr U_{t}\psi_1,U_{t}\psi_2\kkt_{\eta_+}
    =(4\mu\lambda)^{-1} \bbr \Psi_1(t),\Psi_2(t)\kkt_{\eta_+}.
    \label{inn-inv-1}
    \end{equation}
Because $\eta_+$ does not depend on $t$, the inner product
$\bbr\cdot,\cdot\kkt_{\eta_+}$ is invariant under the dynamics
generated by $H$, \cite{p1}. This in turn implies that the
right-hand side of (\ref{inn-inv-1}) should be $t$-independent. In
order to see this, we substitute (\ref{2-comp-psi}) and
(\ref{inn-eta+}) in (\ref{inn-inv-1}) and use (\ref{X}) to derive
    \be
    (\psi_1,\psi_2)=\frac{1}{2\mu}
    \left[\br\psi_1(t)|D^{1/2}\psi_2(t)\kt+
    \br\dot\psi_1(t)|D^{-1/2}\dot\psi_2(t)\kt\right].
    \label{inn-inv-2}
    \end{equation}
We can use (\ref{kgt}) to check that the $t$-derivative of the
right-hand side of (\ref{inn-inv-2}) vanishes identically.
Therefore, (\ref{inn-inv-2}) provides a well-defined inner product
on ${\cal V}$. Endowing ${\cal V}$ with this inner product and
(Cauchy) completing the resulting inner product space one obtains
a separable Hilbert space which we shall denote by ${\cal H}$.
This is the physical Hilbert space of the relativistic QM of
Klein-Gordon fields.

The inner product~(\ref{inn-inv-2}) is an example of the invariant
inner products constructed in \cite{cqg}, and as explained there
it has the following appealing properties: 1) It is not only
positive-definite but relativistically invariant\footnote{A
manifestly covariant expression for this inner product is given in
\cite{p59}.}; 2) In the nonrelativistic limit, it tends to the
usual $L^2$-inner product of $L^2(\R^3)$; 3) It coincides with the
inner product obtained in \cite{woodard} within the framework of
constraint quantization\footnote{It is also identical with the
inner product given in \cite{gt1}. This will be clear after we
discuss the connection to the approach of \cite{gt1} in
Section~4.} ; 4) Its restriction to the subspace of
positive-frequency Klein-Gordon fields is identical with the
restriction of the indefinite Klein-Gordon inner product to this
subspace.

As seen from (\ref{inn-inv-1}), the operator $U_t$ for any value
of $t\in\R$ is a unitary operator mapping ${\cal H}$ to the
Hilbert space ${\cal K}$. Let $t_0\in\R$ be an arbitrary initial
time, and $h:{\cal H}\to{\cal H}$ be defined by
    \be
    h:=U_{t_0}^{-1}\,H\,U_{t_0}.
    \label{h}
    \end{equation}
Then, using (\ref{2-comp-psi}), (\ref{2-comp-H}) and
(\ref{U-zero}), one can show that for all $\psi\in{\cal V}$,
    \be
    h\psi=i\hbar\dot\psi,
    \label{h=}
    \end{equation}
where $\dot\psi$ is the element of ${\cal V}$ defined by
    \[ \dot\psi(t):=\frac{d}{dt}\,\psi(t),
    ~~~~~~~~~~~~~\forall
    t\in\R.\]
It is important to note that (\ref{h=}) is not a time-dependent
Schr\"odinger equation determining $t$-dependence of $\psi(t)$. It
is rather the definition of the operator $h$.

The time-evolution generated by $h$ via the Schr\"odinger equation
    \be
    i\hbar\frac{d}{dt}\psi_t=h\psi_t
    \label{sch-eq-h}
    \end{equation}
is precisely the time-translations in the space ${\cal V}$ of the
solutions of the Klein-Gordon equation~(\ref{kgt}), i.e., if
$\psi_0=\psi_{t_0}$ is the initial value for the one-parameter
family of elements $\psi_t$ of ${\cal V}$, then for all
$t,t'\in\R$, $\psi_t(t')=(e^{-i(t-t_0)h/\hbar}\psi_0)(t')=
\psi_0(t'+t-t_0)$.\footnote{For a derivation of this relation and
a detailed discussion of the difference between (\ref{h=}) and
(\ref{sch-eq-h}) see Sec.~4.2 of \cite{p9}.}

 Furthermore, using the fact that $U_{t_0}$ is a
unitary operator and that $H$ is Hermitian with respect to the
inner product (\ref{inn-eta+}) of ${\cal K}$, we can infer that
$h$ is Hermitian with respect to the inner
product~(\ref{inn-inv-1}) of ${\cal H}$, i.e., time-translations
correspond to unitary transformations of the physical Hilbert
space ${\cal H}$.

Next, we respectively define ${\cal U}:{\cal H}\to{\cal H}'$ and
$H':{\cal H}'\to{\cal H}'$ by
    \bea
    {\cal U}&=&\rho\: U_{t_0}
    \label{U=}\\
    H'&:=&{\cal U}\: h \:{\cal U}^{-1}=\rho H\rho^{-1},
    \label{H-prime}
    \eea
where $\rho$ is the unique positive square root of $\eta_+$. It is
not difficult to see that
    \be
    \rho=\frac{1}{2}\,\left(\begin{array}{cc}
    X+X^{-1} & X-X^{-1}\\
    X-X^{-1} & X+X^{-1}\end{array}\right),~~~~~~
    \rho^{-1}=\frac{1}{2}\,\left(\begin{array}{cc}
    X^{-1}+X & X^{-1}-X\\
    X^{-1}-X & X^{-1}+X\end{array}\right).
    \label{rho}
    \end{equation}
We can check that the operator $\rho$ viewed as mapping ${\cal K}$
onto ${\cal H'}$ is a unitary transformation; using
$\rho^\dagger=\rho=\sqrt\eta_+$, we have
$\br\rho\,\xi,\rho\,\zeta\kt=\bbr\xi,\zeta\kkt_{\eta_+}$ for all
$\xi,\zeta\in{\cal K}$. This in turn implies that ${\cal U}:{\cal
H}\to{\cal H}'$ is also a unitary transformation,
    \be
    \br{\cal U}\psi_1,{\cal U}\psi_2\kt=(\psi_1,\psi_2),
    ~~~~~~~~~~~\forall\psi_1,\psi_2\in{\cal H},
    \label{unitary-u}
    \end{equation}
and that $H'$ must be a Hermitian Hamiltonian operator acting in
${\cal H}'$.

We can compute $H'$ by substituting (\ref{rho}) and
(\ref{2-comp-H}) in (\ref{H-prime}). This leads to the remarkable
result:
    \be
    H'=\hbar\,\left(\begin{array}{cc}
    \sqrt D & 0\\
    0 & -\sqrt D \end{array}\right)=\hbar\sqrt D\,\sigma_3,
    \label{foldy}
    \end{equation}
which is manifestly Hermitian with respect to the inner product
$\br\cdot,\cdot\kt$ of ${\cal H}'$. The Hamiltonian $H'$ is
precisely the Foldy Hamiltonian~\cite{foldy}. Here it is obtained
in an attempt to devise a QM of Klein-Gordon fields that allows
for a genuine probabilistic interpretation without restricting to
the subspace of positive-frequency fields.

Next, we compute the explicit form of the unitary operator ${\cal
U}$ and its inverse. Using (\ref{U=}), (\ref{rho}), (\ref{X}), and
(\ref{U-zero}), we have for all $\psi\in{\cal H}$,
    \be
    {\cal U}\psi=\frac{1}{2\sqrt{\mu}}\,\left(\begin{array}{c}
    D^{1/4}\psi(t_0)+iD^{-1/4}\dot\psi(t_0)\\
    D^{1/4}\psi(t_0)-iD^{-1/4}\dot\psi(t_0)
    \end{array}\right).
    \label{u-psi=}
    \end{equation}
The fact that the arbitrary parameter $\lambda$, introduced in the
two-component formulation of the Klein-Gordon equation, does not
appear in (\ref{foldy}) and (\ref{u-psi=}) is remarkable. The
inverse of ${\cal U}^{-1}$ is also easy to calculate. Let
$\xi\in{\cal H}'$ be a two-component vector (as in
(\ref{xi-zeta})) with components $\xi^1$ and $\xi^2$ belonging to
the domain ${\cal D}$ of $D$. Then in view of (\ref{u-psi=}),
${\cal U}^{-1}\xi$ is the Klein-Gordon field $\psi\in{\cal H}$
satisfying the following initial conditions.
    \be
    \psi(t_0)=\sqrt\mu\,D^{-1/4}(\xi^1+\xi^2),
    ~~~~~~~~~~~
    \dot\psi(t_0)=-i\sqrt\mu\,D^{1/4}(\xi^1-\xi^2).
    \label{U-inverse}
    \end{equation}
Next, we recall that for any Klein-Gordon field $\psi$ and
$t\in\R$, we can express $\psi(t)$ in terms of the initial data
$(\psi(t_0),\dot\psi(t_0))$ according to \cite{fulling,p9}
    \be
    \psi(t)=\cos[(t-t_0)D^{1/2}]\psi(t_0)+
    \sin[(t-t_0)D^{1/2}]D^{-1/2}\dot\psi(t_0).
    \label{psi=ini}
    \end{equation}
Combining (\ref{U-inverse}) and (\ref{psi=ini}), we find
    \be
    [{\cal U}^{-1}\xi](t)=\sqrt\mu\,D^{-1/4}\left[
    e^{-i(t-t_0)D^{1/2}}\xi^1+e^{i(t-t_0)D^{1/2}}\xi^2\right].
    \label{U-inverse-2}
    \end{equation}

In light of the above analysis, the pairs $({\cal H},h)$, $({\cal
K},H)$, and $({\cal H}',H')$ are mutually unitarily equivalent;
they represent the same quantum system. In particular, we can
identify the observables of this quantum system and explore its
symmetries using any of these pairs.

\section{${\cal PT}$-, ${\cal C}$-, and ${\cal CPT}$-Symmetries
of Klein-Gordon Fields}

According to \cite{p7}, the generalized parity ${\cal P}$,
time-reversal ${\cal T}$, and charge-conjugation ${\cal C}$ for a
quasi-Hermitian Hamiltonian with a nondegenerate discrete spectrum
are given by
    \bea
    {\cal P}&=&\sum_n (-1)^n |\phi_n\kt\br\phi_n|,
    \label{P}\\
    {\cal T}&=&\sum_n (-1)^n |\phi_n\kt~\star~\br\phi_n|,
    \label{T}\\
    {\cal C}&=&\sum_n (-1)^n |\psi_n\kt \br\phi_n|,
    \label{C}
    \eea
where $n$ is a spectral label taking nonnegative integer values,
$\{\psi_n,\phi_n\}$ is a complete biorthonormal system with
$\psi_n$ and $\phi_n$ being respectively the eigenvectors of $H$
and $H^\dagger$, $\star$ is the complex-conjugation operator
defined by, for all complex numbers $z$ and state vectors
$\psi,\phi$,
    \[ \star z=z^*,~~~~~~~~~~~~~
    (\star\br\phi|)|\psi\kt:=\star\,\br\phi|\psi\kt=
    \br\psi|\phi\kt,\]
and the positive-definite metric has the form
    \be
    \eta_+=\sum_n |\phi_n\kt\br\phi_n|.
    \label{eta+=}
    \end{equation}
Note that as the labelling of the eigenvectors is arbitrary, so is
the assignment of the signs $(-1)^n$ in (\ref{P}) -- (\ref{C}).

For the Hamiltonian (\ref{2-comp-H}), there is a natural choice of
a sign assignment for the eigenvectors. This is associated with
the label $\epsilon$ appearing in the expression for the
eigenvalues and eigenvectors of $H$. Using this sign assignment
and (\ref{P}) -- (\ref{C}), we obtain the following generalized
parity,  time-reversal, and charge-conjugation operators for the
two-component Klein-Gordon fields.
    \bea
    {\cal P}&:=&\sum_{\epsilon=\pm}\int_{\R^3}d^3k~\epsilon
    |\Phi_{\epsilon,\vec k}\kt\br\Phi_{\epsilon,\vec k}|,
    \label{P-kg}\\
    {\cal T}&:=&\sum_{\epsilon=\pm}\int_{\R^3}d^3k~\epsilon
    |\Phi_{\epsilon,\vec k}\kt\,\star\,\br\Phi_{\epsilon,\vec k}|,
    \label{T-kg}\\
    {\cal C}&:=&\sum_{\epsilon=\pm}\int_{\R^3}d^3k~\epsilon
    |\Psi_{\epsilon,\vec k}\kt\br\Phi_{\epsilon,\vec k}|.
    \label{C-kg}
    \eea

Substituting (\ref{eigenvector-H}) and
(\ref{eigenvector-H-dagger}) in these equations and doing the
necessary algebra, we find the rather remarkable result:
    \be
    {\cal P}=\sigma_3,~~~~~~~~~~~{\cal T}=\sigma_3\star,
    ~~~~~~~~~~~~{\cal PT}=\star.
    \label{P-T-PT}
    \end{equation}
Hence, the ${\cal PT}$-symmetry of the
Hamiltonian~(\ref{2-comp-H}) is equivalent to the statement that
it is a real operator. Similarly, we compute
    \be
    {\cal C}=\frac{1}{2}\left(\begin{array}{cc}
    X^2+X^{-2} & X^2-X^{-2}\\
    -X^2+X^{-2} & -(X^2+X^{-2})\end{array}\right),
    \label{C-kg=}
    \end{equation}
which in view of (\ref{eta+}) is consistent with the identity
${\cal C}=\eta_+^{-1}{\cal P}$ (equivalently $\eta_+={\cal PC}$.)

By construction, ${\cal C}$ generates a symmetry of the
Hamiltonian (\ref{2-comp-H}). The meaning of this symmetry becomes
clear, once we use the following alternative expression for ${\cal
C}$:
    \be
    {\cal C}=\hbar^{-1}D^{-1/2}H=\frac{H}{\sqrt{H^2}}.
    \label{C-kg=H}
    \end{equation}
Here we have made used of (\ref{C-kg=}), (\ref{X}),
(\ref{2-comp-H}) and the fact that unlike $H$, $H^2$ is a
positive-definite operator acting in ${\cal H}'$. This follows
from the identity $H^2=\hbar^2 D I$, where $I$ is the $2\times 2$
identity matrix. According to (\ref{C-kg=H}), ${\cal C}$ is a
$\Z_2$-grading operator for the Hilbert space that splits it into
the spans of the eigenvectors of $H$ with positive and negative
eigenvalues, respectively.

We can use the unitary operator $U_{t_0}:{\cal H}\to{\cal K}$ to
define the generalized parity, time-reversal, and
charge-conjugation operators for the ordinary single-component
Klein-Gordon fields. These are given by
    \be
    {\rm P}:=U_{t_0}^{-1}{\cal P}U_{t_0},~~~~~~~~~~
    {\rm T}:=U_{t_0}^{-1}{\cal T}U_{t_0},~~~~~~~~~~
    {\rm C}:=U_{t_0}^{-1}{\cal C}U_{t_0}.
    \label{P-T-C}
    \end{equation}
Using the definition of the operator $U_{t_0}$, we may obtain
expressions for the action of the operators ${\rm P}$ and ${\rm
T}$ on a given Klein-Gordon field $\psi\in{\cal H}$. More
interesting are the corresponding expressions for the action of
${\rm PT}$ and ${\rm C}$, that actually generate symmetries of the
Hamiltonian $h$. It turns out that, for all $\psi\in{\cal H}$ and
$t\in\R$,
    \be
    ({\rm PT}\psi)(t)=\psi(-t).
    \label{time-rev}
    \end{equation}
Hence ${\rm PT}$ is just the ordinary time-reversal operator, and
{\em the ${\rm PT}$-symmetry of $h$ means that the order in which
one performs a time-translation and a time-reversal transformation
on a Klein-Gordon field is not important.}

Next, consider the subspaces ${\cal V}_+$ and ${\cal V}_-$ of
${\cal V}$ that are respectively spanned by the positive- and
negative-energy eigenvectors of $h$. The elements $\psi_\pm$ of
${\cal V}_\pm$ are of the form
    \be
    \psi_\pm(t)=\int_{\R^3}d^3k~
    e^{\mp i\omega_kt}f(\vec k)\:\phi_{\vec k}
    \label{psi-pm}
    \end{equation}
for some $f\in L^2(\R^3)$. Clearly ${\cal V}={\cal V}_+\oplus
{\cal V}_-$, and restricting the inner product $(\cdot,\cdot)$ to
${\cal V}_\pm$ (and completing the resulting inner product spaces)
we obtain Hilbert subspaces ${\cal H}_\pm$ of ${\cal H}$. In view
of (\ref{inn-inv-2}) and (\ref{psi-pm}), we can further show that
    \[(\psi_+,\psi_-)=0,~~~~~~~~~~~\forall \psi_\pm\in{\cal
    V}_\pm.\]
This is sufficient to infer that indeed ${\cal H}={\cal
H}_+\oplus{\cal H}_-$. The generalized charge-conjugation operator
${\rm C}$ defined by (\ref{P-T-C}) is actually the grading
operator associated with this orthogonal direct sum decomposition
of ${\cal H}$, i.e., if $\psi=\psi_++\psi_-$ such that
$\psi_\pm\in{\cal H}_\pm$, then
    \be
    {\rm C}\,\psi=\psi_+-\psi_-.
    \label{polarization}
    \end{equation}
In other words, ${\rm C}$ is the operator that decomposes the
Hilbert space into its positive- and negative-energy subspaces. In
view of the fact that for a complex Klein-Gordon field the
positive and negative energies respectively occur for positive and
negative charges,  ${\rm C}$ is identical with the charge-grading
operator. As a result, {\em the ${\rm C}$-symmetry of $h$ means
that the charge of a Klein-Gordon field does not change sign under
time-translations}, and {\em the ${\rm CPT}$-symmetry of $h$ is
equivalent to the statement that the combined action of
time-reversal and charge-grading operators commutes with any
time-translation of a Klein-Gordon field.}

Next, we use the unitary operator $\rho:{\cal K}\to{\cal H}'$ to
express the generalized parity, time-reversal, and
charge-conjugation operators in the Foldy representation of the
Klein-Gordon fields which is based on the Hilbert space ${\cal
H}'$ and the Hamiltonian $H'$. They are given by
    \be
    {\cal P}':=\rho\,{\cal P}\rho^{-1},~~~~~~~~~~
    {\cal T}':=\rho\,{\cal T}\rho^{-1},~~~~~~~~~~
    {\cal C}':=\rho\,{\cal C}\rho^{-1}.
    \label{PTC-prime}
    \end{equation}
Again, we compute the form of the symmetry generators ${\cal
P}'{\cal T}'$ and ${\cal C}'$. Because $\rho$ and $\rho^{-1}$ are
real operators,
    \be
    {\cal P}'{\cal T}'={\cal PT}=\star.
    \label{PT-prime}
    \end{equation}
Moreover, using (\ref{C-kg=H}), (\ref{H-prime}) and
(\ref{PTC-prime}),
    \be
    {\cal C}'=\frac{H'}{\sqrt{H^{'2}}}=\sigma_3.
    \label{C-prime=}
    \end{equation}
Clearly, the ${\cal P}'{\cal T}'$-symmetry of $H'$ is related to
the fact that $H'$ is a real operator, and the ${\cal
C}'$-symmetry of $H'$ is because it is proportional to ${\cal
C}'$. Obviously, the physical interpretation of the generalized
$PT$- and $C$-symmetries is independent of the choice of the
unitary-equivalent representations of the underlying quantum
system. As seen from the above analysis, the representation based
on the Hilbert space ${\cal H}$ and Hamiltonian $h$ is useful in
identifying ${\cal PT}$ as ordinary time-reversal operator, while
the Foldy representation is useful in identifying ${\cal C}$ with
the charge-grading operator.

\section{Physical Observables and Wave Functions for Klein-Gordon
Fields}

In the preceding section we introduced three equivalent
representations of the QM of a Klein-Gordon field. These
corresponded to the following choices for the pair (Hilbert space,
Hamiltonian): $({\cal H},h)$,~$({\cal K},H)$,~and $({\cal
H}',H')$. We can study the physical observables of this quantum
theory using any of these representations. We will employ the
usual notion of quantum mechanical observables, namely identify
them with Hermitian operators acting in the Hilbert space. Because
${\cal H}'=L^2(\R^3)\oplus L^2(\R^3)$, the Foldy representation
$({\cal H}',H')$ is more convenient for the construction of the
observables. Once this is done, we may use the unitary map ${\cal
U}:{\cal H}\to{\cal H}'$ to obtain the form of the observables in
the representation $({\cal H},h)$.

First, we introduce the following set of basic observables (in the
Foldy representation) that can be used to construct others.
    \be
    \vec X_\mu:=\vec{\rm x}\otimes\sigma_\mu,~~~~~~
    \vec P_\mu:=\vec{\rm p}\otimes\sigma_\mu,~~~~~~
    S_\mu:=1\otimes\sigma_\mu.
    \label{ob-0}
    \end{equation}
Here, $\mu\in\{0,1,2,3\}$, $\vec{\rm x}$, $\vec{\rm p}$, and $1$
are the position, momentum, and identity operators acting in
$L^2(\R^3)$, $\sigma_0=I$ is the $2\times 2$ identity matrix, and
$\sigma_\mu$ with $\mu\neq 0$ are the Pauli matrices:
$\sigma_1=${\tiny$\left(\begin{array}{cc}
    0 & 1 \\
    1 & 0\end{array}\right)$},
    $\sigma_1=${\tiny$\left(\begin{array}{cc}
    0 & -i \\
    i & 0\end{array}\right)$},
    $\sigma_3=${\tiny$\left(\begin{array}{cc}
    1 & 0 \\
    0 & -1\end{array}\right)$}.
In the following, we will adopt the abbreviated notation of not
writing `$1\otimes$' explicitly. In particular, we will identify
$S_\mu$ with $\sigma_\mu$.

Clearly, $\vec X_0$ and $\sigma_3$ form a maximal set of commuting
operators. In particular, we can use their common `eigenvectors'
($\xi_{\epsilon,\vec x}$) to construct a basis of ${\cal H}'$.
These are defined, for all $\vec x\in\R^3$ and
$\epsilon\in\{-,+\}$, by
    \be
    \xi_{\epsilon,\vec x}:=|\vec x\kt\otimes e_\epsilon,
    \label{xi}
    \end{equation}
where $e_+=${\tiny $\left(\begin{array}{c} 1 \\
0\end{array}\right)$}, $e_-=${\tiny $\left(\begin{array}{c} 0 \\
1\end{array}\right)$}, and $|\vec x\kt$ are the $\delta$-function
normalized position kets satisfying
    \be
    \vec{\rm x}|\vec x\kt=\vec x|\vec x\kt,~~~~~~~~
    \br x|x'\kt=\delta^3(\vec x-\vec x'),~~~~~~~~
    \int_{\R^3}d^3x\:|\vec x\kt\br\vec x|=1.
    \label{x-ket}
    \end{equation}
It is easy to see that indeed $\vec X_0\,\xi_{\epsilon,\vec
x}=\vec x\,\xi_{\epsilon,\vec x}$ and
$\sigma_3\,\xi_{\epsilon,\vec x}=\epsilon\,\xi_{\epsilon,\vec x}$
. Furthermore,
    \be
    \br \xi_{\epsilon,\vec x},\xi_{\vec x',\epsilon'}\kt=
    \delta_{\epsilon,\epsilon'}\delta^3(\vec x-\vec x'),~~~~~~~~~~~
    \sum_{\epsilon=\pm}\int_{\R^3}d^3x\:|\xi_{\epsilon,\vec x}\kt
    \br\xi_{\epsilon,\vec x}|=\sigma_0.
    \label{orthonormal}
    \end{equation}

We can express any two-component vector $\Psi'\in{\cal H}'$ in the
basis $\{\xi_{\epsilon,\vec x}\}$ according to
    \be
    \Psi'=\sum_{\epsilon=\pm}\int_{\R^3}d^3x\:
    f(\epsilon,\vec x)\,\xi_{\epsilon,\vec x},
    \label{wf-1}
    \end{equation}
where $f:\{-,+\}\times\R^3\to\C$ is the wave function associated
with $\Psi'$ in the position-representation, i.e.,
    \be
    f(\epsilon,\vec x):=\br\xi_{\epsilon,\vec x},\Psi'\kt.
    \label{wf-0}
    \end{equation}
As is well-known from nonrelativistic QM, one can also express the
observables as linear operators acting on the wave functions $f$.
For example, let $O':{\cal H}'\to{\cal H}'$ be a Hermitian
operator defining a physical observables in the Foldy
representation and $\Psi'={\cal U}\psi$ describe a Klein-Gordon
field $\psi\in{\cal H}$ in this representation. Then
    \be
    O'\Psi'=\sum_{\epsilon=\pm}\int_{\R^3}d^3x\:
    [\hat O f(\epsilon,\vec x)]\,\xi_{\epsilon,\vec x},
    \label{ob-1}
    \end{equation}
where
    \bea
    \hat O f(\epsilon,\vec x)&:=&
    \sum_{\epsilon'=\pm}\int_{\R^3}d^3x'
    \hat O(\epsilon,\vec x;\epsilon',\vec x')f(\epsilon',\vec x'),
    \label{ob-2}\\
    \hat O(\epsilon,\vec x;\epsilon',\vec x')&:=&
    \br\xi_{\epsilon,\vec x},O'\xi_{\epsilon',\vec x'}\kt.
    \label{ob-3}
    \eea

Next, we introduce the operators
    \be
    \vec x_\mu:={\cal U}^{-1}\vec X_\mu {\cal U},~~~~~~~~~~~~~~
    \vec p_\mu:={\cal U}^{-1}\vec P_\mu {\cal U},~~~~~~~~~~~~~~
    s_\mu:={\cal U}^{-1}\sigma_\mu {\cal U},
    \label{ob-4}
    \end{equation}
that act in ${\cal H}$, and define the Klein-Gordon fields
    \be
    \psi_{\epsilon,\vec x}:={\cal U}^{-1}\xi_{\epsilon,\vec x}.
    \label{local}
    \end{equation}
Clearly, (\ref{ob-4}) describe the same physical observables as
(\ref{ob-0}), albeit in the representation $({\cal H},h)$. The
fields~(\ref{local}) form a complete orthonormal basis of ${\cal
H}$; one can easily check using (\ref{unitary-u}),
(\ref{orthonormal}), and (\ref{local}) that
    \be
    (\psi_{\epsilon,\vec x},\psi_{\epsilon',\vec x'})=
    \delta_{\epsilon,\epsilon'} \delta^3(\vec x-\vec x'),~~~~~~~~~~~
    \sum_{\epsilon=\pm}\int_{\R^3}d^3x\:|\psi_{\epsilon,\vec x})
    (\psi_{\epsilon,\vec x}|=s_0.
    \label{orthonormal-psi}
    \end{equation}
Note that here for all $\psi_1,\psi_2\in{\cal H}$, the operator
$|\psi_1)(\psi_2|$ is defined by
$|\psi_1)(\psi_2|\psi_3:=(\psi_2,\psi_3)\psi_1$, for any
$\psi_3\in{\cal H}$, and that $s_0$ coincides with the identity
operator for ${\cal H}$.

Again, any Klein-Gordon field $\psi$ may be expressed in the basis
$\{\psi_{\epsilon,\vec x}\}$ in terms of the wave
functions~(\ref{wf-0}) according to
    \be
    \psi=\sum_{\epsilon=\pm}\int_{\R^3}d^3x\:
    f(\epsilon,\vec x)\,\psi_{\epsilon,\vec x}.
    \label{wf}
    \end{equation}
This follows from (\ref{orthonormal-psi}) and
$(\psi_{\epsilon,\vec x},\psi)=
    \br{\cal U}\psi_{\epsilon,\vec x},
    {\cal U}\psi\kt=\br\xi_{\epsilon,\vec x},\Psi'\kt=
    f(\epsilon,\vec x).$
Similarly, any Hermitian operator $o:{\cal H}\to{\cal H}$
associated with a physical observable satisfies
    \be
    o\,\psi=\sum_{\epsilon=\pm}\int_{\R^3}d^3x
    [\hat O f(\epsilon,\vec x)]\psi_{\epsilon,\vec x},
    \label{ob-5}
    \end{equation}
where $\hat O f(\epsilon,\vec x)$ is defined by (\ref{ob-2}) and
    \be
    \hat O(\epsilon,\vec x;\epsilon',\vec x'):=
    (\psi_{\epsilon,\vec x},o\,\psi_{\epsilon',\vec x'}).
    \label{ob-6}
    \end{equation}

The wave functions $f$ also provide a description of the quantum
system associated with the Klein-Gordon fields. To see this, first
we use (\ref{wf}) and (\ref{orthonormal-psi}) to compute the inner
product of a pair of Klein-Gordon fields, $\psi,\gamma\in{\cal
H}$, in terms of their wave functions  $f(\epsilon,\vec
x):=(\psi_{\epsilon,\vec x},\psi)$ and $g(\epsilon,\vec
x):=(\psi_{\epsilon,\vec x},\gamma)$. This yields
    \[ (\psi,\gamma)=\sum_{\epsilon=\pm}\int_{\R^3}d^3x\:
        f(\epsilon,\vec x)^*g(\epsilon,\vec x).\]
More generally, for any Hermitian operator $o:{\cal H}\to{\cal H}$
describing an observable, we have
    \be
    (\psi,o\,\gamma)=\sum_{\epsilon=\pm}\int_{\R^3}d^3x\:
        f(\epsilon,\vec x)^*\, \hat O g(\epsilon,\vec x)=
        \sum_{\epsilon=\pm}\int_{\R^3}d^3x\:
        [\hat O f(\epsilon,\vec x)]^* g(\epsilon,\vec x).
    \label{ob-7}
    \end{equation}
This shows that the wave functions $f$ may be viewed as elements
of ${\cal H}'$, and the observables may be described by Hermitian
operators $\hat O$ acting on the wave functions. For example, the
action of $\vec x_0$, $\vec p_0$, and $s_3$ on $\psi$ corresponds
to the action of the operators $\hat{\vec x}_0$, $\hat{\vec p}_0$,
and $\hat s_3$ on $f$, where
    \be
    \hat{\vec x}_0f(\epsilon,\vec x)=\vec x f(\epsilon,\vec x),
    ~~~~~~~~~~~
    \hat{\vec p}_0f(\epsilon,\vec x)=-i\hbar\vec\nabla
    f(\epsilon,\vec x),~~~~~~~~~~~
    \hat s_3 f(\epsilon,\vec x)=\epsilon f(\epsilon,\vec x).
    \label{wf-ob}
    \end{equation}
Similarly, the action of the Hamiltonian $h$ on $\psi$ corresponds
to the action,
    \be
    \hat h f(\epsilon,\vec x)=\hbar\epsilon\sqrt{-\nabla^2+\mu^2}
    f(\epsilon,\vec x),
    \label{h-hat}
    \end{equation}
of the operator
    \be
    \hat h:=\hat s_3\sqrt{\hat{\vec p}_0^2+m^2}
    \label{h-hat=}
    \end{equation}
on the wave function $f$.

Having expressed $h$ in terms of the wave functions $f$, we can
also obtain the explicit form of the Schr\"odinger
equation~(\ref{sch-eq-h}) as a partial differential equation for
$f$. The result is
    \be
    i\hbar\partial_t f(\epsilon,\vec x;t)=
    \epsilon\sqrt{-\hbar^2\nabla^2+m^2}\; f(\epsilon,\vec x;t),
    \label{sch-eq-f}
    \end{equation}
where $f(\epsilon,\vec x;t)=(\psi_{\epsilon,\vec x},\psi_t)$ is
the wave function for the one-parameter family of the
(time-translated) Klein-Gordon fields $\psi_t$. Furthermore,
applying $i\partial_t$ to both side (\ref{sch-eq-f}), we can check
that the wave functions $f$ also satisfies the Klein-Gordon
equation:
    \be
    [\partial_t^2-\nabla^2+\mu^2]f(\epsilon,\vec x;t)=0.
    \label{kg-f}
    \end{equation}
We wish to emphasize that we have obtained this equation by
iterating~the Schr\"odinger equation~(\ref{sch-eq-f}). Unlike the
Klein-Gordon equation, the solutions of (\ref{sch-eq-f}) are
uniquely determined by a single initial condition.

Next, recall that because the time-reversal operator
(\ref{PT-prime}) acting in ${\cal H}'$ commutes with $\vec X_0$,
the eigenvectors $\xi_{\epsilon,\vec x}$ may be taken to be real.
In this case the action of the time-reversal operator $T={\rm PT}$
on any $\psi\in{\cal H}$ is equivalent to the complex-conjugation
of the associated wave-function $f$. Denoting by $\hat T$ the
time-reversal operator acting on $f$, we have $\hat T
f(\epsilon,\vec x)=f(\epsilon,\vec x)^*$. Similarly, we can
identify the operator $\hat s_3$ with the charge-grading operator
acting on the wave functions.

The description of the QM of the Klein-Gordon fields using the
wave functions $f(\epsilon,\vec x)$ is identical with the one
offered in \cite{gt1}. The two approaches differ in that ours is
motivated and relies on the construction of a positive-definite,
relativistic invariant, and conserved inner product on the
solution space of the Klein-Gordon equation whereas that of
\cite{gt1} rests on the canonical quantization of the associated
classical relativistic particle and obtaining an appropriate
representation of the algebra of basic observables. The fact that
both approaches lead to unitary-equivalent pictures is quite
remarkable and most comforting.

As we shall see in the following section $f(\epsilon,\vec x)$ is
the position wave function for a state vector (field)
$\psi\in{\cal H}$ where the position operator is precisely the
Newton-Wigner operator \cite{newton-wigner} (for $\epsilon=+1$ and
its analog for $\epsilon=-1$). The latter is well-known not to be
relativistically covariant. Hence the description of the system
using the wave functions $f(\epsilon,\vec x)$ is a non-covariant
description of a quantum system that does admit a
unitary-equivalent covariant description, namely the one offered
by the Hilbert space ${\cal H}$ and the Hamiltonian $h$.

\section{Position Operators and Localized States}

In canonical approach to QM, the observables of a quantum system
are described by Hermitian operators acting in the Hilbert space.
The physical interpretation of these operators, however, rests on
the quantization scheme, i.e., the way in which the Hermitian
operators are related to the classical observables. The classical
system associated with the Klein-Gordon fields is a relativistic
free particle whose energy is given by
$E=\pm\sqrt{p^2+m^2}$.\footnote{Note that the negative sign is not
ruled out by special relativity.} Performing canonical
quantization \cite{dirac-qm} on this system, i.e., setting $\vec
p\to-i\hbar\vec\nabla$, one finds
$E\to\pm\hbar\sqrt{-\nabla^2+\mu^2}$. In view of (\ref{h-hat}) and
(\ref{h-hat=}), this implies that the canonical quantization is
relevant to the description of the Klein-Gordon fields $\psi$ in
terms of their wave functions $f$.\footnote{This is actually the
case as shown in \cite{gt1}.} As a result, the operators
$\hat{\vec x}_0$ and $\hat{\vec p}_0$ that clearly satisfy the
canonical commutation relations may be identified with the
position and momentum operators acting in the space of the wave
functions $f$. This in turn means that the operators $\vec X_0$
and $\vec P_0$ in the Foldy representation and the operators $\vec
x_0$ and $\vec p_0$ in the $({\cal H},h)$-representation also
describe the position and momentum observables. In particular, the
basis vectors $\xi_{\epsilon,\vec x}$ and $\psi_{\epsilon,\vec x}$
determine the states of the system with a definite position value
$\vec x$; they are {\em localized} in space. They also have
definite charge.

Next, we obtain the explicit form of the position operator $\vec
x_0$ that is defined to act on the Klein-Gordon fields
$\psi\in{\cal H}$. Note that $\vec\chi:=\vec x_0\psi$ is a
three-component field whose components satisfy the Klein-Gordon
equation~(\ref{kgt}). It is uniquely determined in terms of its
initial data $(\vec\chi(t_0),\dot{\vec\chi}(t_0))$ for some
$t_0\in\R$. We can compute the latter using (\ref{ob-4}),
(\ref{u-psi=}), (\ref{U-inverse}), and the identities:
$D=\hbar^{-2}(\vec{\rm p}^2+m^2)$ and $[F(\vec{\rm p}),\vec{\rm
x}]=-i\hbar\vec\nabla F(\vec{\rm p})$, where $F$ is a
differentiable function. This yields
    \be
    \vec \chi(t_0)=\vec{\cal X}\psi(t_0),~~~~~~~~~~~
    \dot{\vec\chi}(t_0)=\vec{\cal X}^\dagger\dot\psi(t_0),
    \label{x-psi}
    \end{equation}
where
    \be
    \vec{\cal X}:=\vec{\rm x}+\frac{i\hbar\,\vec{\rm p}}{2(\vec
    {\rm p}^2+m^2)}.
    \label{NW}
    \end{equation}

We can employ (\ref{psi=ini}) to express $\vec\chi(t)$ in terms of
(\ref{x-psi}). The resulting expression is
    \be
    \vec\chi(t)=\cos[(t-t_0)D^{1/2}]\vec\chi(t_0)+
    \sin[(t-t_0)D^{1/2}]D^{-1/2}\dot{\vec\chi}(t_0),
    ~~~~~~~~~~\forall t\in\R.
    \label{chi=}
    \end{equation}
Now, substituting (\ref{x-psi}) and (\ref{NW}) in this equation,
using the power series expansion of $\sin$ and $\cos$, and doing
the necessary calculations, we find
    \be
    [\vec x_0\psi](t)=\vec\chi(t)=\vec{\rm x}\,\psi(t)-\vec q\,\left[J_1(t-t_0)\psi(t_0)+
    J_2(t-t_0)\dot\psi(t_0)\right],~~~~~~~~~~\forall t\in\R,
    \label{chi=2}
    \end{equation}
where
    \be
    \vec q:=\frac{i\hbar\,\vec{\rm p}}{2(\vec
    {\rm p}^2+m^2)},
    \label{q=}
    \end{equation}
and for all $\tau\in\R$
    \bea
    J_1(\tau)&:=&-\cos(\tau D^{1/2})-2\tau
    \sin(\tau D^{1/2})D^{1/2}=\sum_{\ell=0}^\infty
    \left[\frac{(-1)^\ell
    (4\ell-1)}{(2\ell)!}\right]\tau^{2\ell}D^\ell,\nn\\
    J_2(\tau)&:=&2\tau \cos(\tau D^{1/2})-
    \sin(\tau D^{1/2})D^{-1/2}=
    \sum_{\ell=0}^\infty
    \left[\frac{(-1)^\ell
    (4\ell+1)}{(2\ell+1)!}\right]\tau^{2\ell+1}D^\ell.\nn
    \eea

In addition to being a Hermitian operator acting in the physical
Hilbert space ${\cal H}$, the position operator $\vec x_0$ has the
following notable properties.
    \begin{enumerate}
    \item In view of (\ref{x-psi}) and (\ref{NW}), it coincides
    with the Newton-Wigner position operator
    \cite{newton-wigner,san-good}, if it is restricted to the
    positive-frequency Klein-Gordon fields. Indeed, (\ref{chi=2})
    provides an explicit form of the Newton-Wigner position operator
    that, to the best of our knowledge, has not been previously
    given.
    \item It respects the charge superselection rule
    \cite{superselection,klauder}, for it commutes with the
    charge-grading operator ${\rm C}=s_3$. This is easily seen
    by noting that $\vec x_0$ and $s_3$ are  respectively obtained
    via a similarity transformation (\ref{ob-4}) from
    $\vec X_0$ and $\vec S_3=\sigma_3$, and that according to
    (\ref{ob-0}), $[\vec X_0,\sigma_3]=0$.
    \item It has the correct nonrelativistic limit: as
    $c\to\infty$, $[\vec x_0\psi](t)\to \vec{\rm x}\psi(t)$.
    \end{enumerate}

We can similarly evaluate the action of the momentum operator
$\vec p_0$ on $\psi$. Because $\vec P_0$ and $\rho$ commute, in
view of (\ref{ob-4}) and (\ref{U=}), we have $\vec
p_0=U_{t_0}^{-1}\vec P_0 U_{t_0}$. This in turn implies $[\vec
p\,\psi](t)=\vec{\rm p}\,\psi(t)$ for all $t\in\R$. Furthermore,
in view of the fact that for every differentiable element
$\varphi$ of $L^2(\R^3)$ with nonzero gradient, $\vec q\,\varphi$
is along $\vec{\rm p}\,\varphi$, we can show that the angular
momentum operator $\vec L:=\vec x_0\times\vec p_0$ acts on $\psi$
according to $[\vec{L}\,\psi](t)=\vec{\rm x}\times\vec{\rm
p}\,\psi(t)$ for all $t\in\R$. Therefore, unlike the position
operator $\vec x_0$, the (linear) momentum $\vec p_0$ and the
angular momentum $\vec L$ operators have the same expressions as
in nonrelativistic QM.

Having obtained the position and momentum observables for
Klein-Gordon fields, we can also introduce a set of coherent
states by requiring that they are the eigenstates of the
annihilation operator $\vec a:=\sqrt{\frac{k}{2\hbar}}\,\left(\vec
x_0+ik^{-1}\vec p_0\right)$, where $k=m\omega\in\R$ is the
characteristic oscillator constant. Because both $\vec x_0$ and
$\vec p_0$ commute with the charge-grading operator $s_3$, so does
the annihilation operator $\vec a$. This suggests that we can
introduce coherent states that have a definite charge. The
corresponding state vectors $|\vec z,\epsilon)$ are defined as the
common eigenvectors of $\vec a$ and $s_3$, i.e., $\vec a|\vec
z,\epsilon)=\vec z|\vec z,\epsilon)$ and $s_3|\vec
z,\epsilon)=\epsilon|\vec z,\epsilon)$, where $\vec z\in\C^3$ and
$\epsilon\in\{-,+\}$.  By construction, the coherent states
represented by the vectors $|\vec z,\epsilon)$ are free from the
subtleties associated with the nontrivial charge structure of the
conventional relativistic coherent states \cite{klauder}.

\section{Conclusion}

In this article, we have explored a concrete realization of the
notions of generalized parity $({\cal P})$, time-reversal $({\cal
T})$, and charge-conjugation $({\cal C})$ operators for
Klein-Gordon fields, and showed that the generalized
parity-time-reversal and charge-conjugation symmetries that arise
in pseudo-Hermitian QM have well-known physical meanings in
relativistic QM; they simply mean that the time-translations of a
Klein-Gordon field commute with the time-reversal and
charge-grading operators. In particular, ${\cal C}$ is not the
usual charge-conjugation operator of relativistic QM, for it does
not map a positive-frequency field to a negative-frequency field.
Rather, it is a grading operator that splits the Hilbert space
into the subspaces of positive- and negative-frequency fields.
Being a Hermitian operator acting the Hilbert space of the system
${\cal C}$ is an observable. Its classical analog and its
significance in the study of Klein-Gordon fields were initially
noted in \cite{gt1} and subsequently employed in \cite{gt2,saa}.

The theory of pseudo-Hermitian operators that was initially
developed for the purpose of making sense of $PT$-symmetric QM
provides an invaluable tool for devising a genuine quantum
mechanical description of Klein-Gordon fields. It allows for an
explicit construction of the Hilbert space ${\cal H}$ of the
solutions of the Klein-Gordon equation. It yields a simple
description of the physical observables in terms of Hermitian
operators acting in the Hilbert space ${\cal H}$ or equivalently
${\cal H}'=L^2(\R^3)\oplus L^2(\R^3)$. It leads to the
introduction of a set of wave functions $f$ that uniquely
determine the Klein-Gordon fields $\psi$ and at the same time
satisfy a Schr\"odinger equation.\footnote{This may be viewed as
an interesting link between relativistic and nonrelativistic QM on
the one hand, and in view of the results of \cite{gt1} the link to
one-particle sector of the corresponding quantum field theory on
the other hand.}

The quantum theory of Klein-Gordon fields outlined in this article
shares almost all the properties of ordinary nonrelativistic
quantum systems. It provides a simple construction of a position
operator that fulfills the requirements of the
charge-superselection rule and allows for the identification of a
set of relativistic localized and coherent states. Restricting to
the space of positive-frequency Klein-Gordon field, this position
operator and consequently the corresponding localized states
coincide with those obtained by Newton and Wigner
\cite{newton-wigner} through their axiomatic construction.

The quantum theory developed here is relativistic in nature, for
the inner product of the Hilbert space ${\cal H}$ and consequently
the transition and scattering amplitudes are invariant under
Lorentz transformations. It also has the QM of a free particle as
its nonrelativistic limit. In particular, the Hilbert space, the
Hamiltonian, and the basic position and momentum operators tend to
their well-known nonrelativistic analogues as one takes
$c\to\infty$.\footnote{This is best seen in the description of the
theory based on the wave functions $f$.}

Finally, we note that the approach pursued in this article can be
easily generalized to Klein-Gordon fields interacting with a
time-independent magnetic field. This amounts to a simple
redefinition of the operator $D$ from (\ref{kg-D}) to
$[D\psi(t)](\vec x):=[-(\vec\nabla-\vec A(\vec
x))^2+\mu^2]\psi(t,\vec x)$, where $\vec A$ is a corresponding
vector potential. In view of the general theory developed in
\cite{p9}, one may attempt to treat the case of time-dependent
electromagnetic fields \cite{p59}.

\subsection*{Acknowledgments}
I wish to thank Ali \"Ulger for a useful discussion and Farhad
Zamani for pointing out a terminological error made in the draft
of this article. This work has been supported by the Turkish
Academy of Sciences in the framework of the Young Researcher Award
Program (EA-T$\ddot{\rm U}$BA-GEB$\dot{\rm I}$P/2001-1-1).

%\newpage

\renewcommand{\baselinestretch}{1.2}

\ed